\documentclass[a4paper]{article}

\usepackage{INTERSPEECH2021}

\usepackage{tikz}
\usepackage{hyperref}
\usepackage{booktabs}
\usepackage{enumitem}
\usepackage{algorithm}
\usepackage[noend]{algpseudocode}
\usepackage{amsmath}
\usepackage{multirow}
\usepackage{bbm}

\DeclareMathOperator*{\argmax}{arg\,max}

\title{Semi-FedSER: Semi-supervised Learning for Speech Emotion Recognition On Federated Learning using Multiview Pseudo-Labeling}
\name{Tiantian Feng$^1$, Shrikanth Narayanan$^1$}
\address{
  $^1$Signal Analysis and Interpretation Lab (SAIL), University of Southern California}
\email{tiantiaf@usc.edu, shri@ee.usc.edu}

\begin{document}

\maketitle
\begin{abstract}

\textbf{S}peech \textbf{E}motion \textbf{R}ecognition (\textbf{SER}) application is frequently associated with privacy concerns as it often acquires and transmits speech data at the client-side to remote cloud platforms for further processing. These speech data can reveal not only  speech content and affective information but the speaker's identity, demographic traits, and health status. Federated learning (FL) is a distributed machine learning algorithm that coordinates clients to train a model collaboratively without sharing local data. This algorithm shows enormous potential for SER applications as sharing raw speech or speech features from a user's device is vulnerable to privacy attacks. However, a major challenge in FL is limited availability of high-quality labeled data samples. In this work, we propose a semi-supervised federated learning framework, Semi-FedSER, that utilizes both labeled and unlabeled data samples to address the challenge of limited labeled data samples in FL. We show that our Semi-FedSER can generate desired SER performance even when the local label rate $l=20\%$ using two SER benchmark datasets: IEMOCAP and MSP-Improv.

\end{abstract}
\vspace{1mm}

\footnote{This paper was submitted to Insterspeech 2022 for review.}

\section{Introduction}

Speech emotion recognition (SER) has gained enormous  interest in many diverse applications such as smart virtual assistants \cite{lee2020study}, medical diagnoses \cite{ramakrishnan2013speech,Bone2017SignalProcessingandMachine}, and education \cite{li2007speech}. SER aims to identify target emotional states conveyed in vocal expressions automatically. The SER system typically has three parts: data acquisition, data transfer, and classification \cite{koolagudi2012emotion}. However, privacy is a major challenge that researchers or service providers often need to face before deploying the SER system. The speech signals are typically sensitive and can reveal a significant amount of information about an individual. These information include speech content and speaker's identity, traits (e.g., age, gender), and states (e.g., health status), many of which are deemed sensitive from an application point of view.

Federated learning (FL) is a popular privacy-preserving distributed learning approach that allows clients to train a model collaboratively without sharing their local data  \cite{mcmahan2017communication}. In an FL setting, a central server aggregates model updates from multiple clients during the training process. Each client generates such model updates by locally training a model on the private data available at the client. Furthermore, this machine learning approach reduces information leaks compared to classical centralized machine learning frameworks since personal data does not leave the client. Therefore, this distributed learning paradigm can be a natural choice for developing real-world multiuser SER applications as sharing raw speech or speech features from users' devices is vulnerable to privacy attacks.

Most current works in FL have focused on fully supervised settings where all the input data are with labels. However, high-quality labeled data samples do not often exist in real-life settings, and most data samples are indeed unlabeled. In order to address the limited labeled data samples in an FL setting, prior works have considered using semi-supervised learning (SSL), which utilizes unlabeled samples in addition to a small portion of labeled examples to obtain desired model performance \cite{lin2021semifed, tsouvalas2022privacy}. SemiFed is an SSL framework that integrates supervised and unsupervised training during FL. This approach also generates pseudo-labels by ranking data samples using peer models. In addition, FedMatch \cite{jeong2020federated} uses an inter-device consistency loss to enforce consistency between the pseudo-labeling predictions made across multiple devices. However,  we argue that these methods are unrealistic in practice for two reasons: first, SemiFed significantly increases the communication costs occurring at each global training round, where the server needs to transfer additional models to each local client; additionally, sharing the local model to peers may cause potential breaches since the prior work has shown that data reconstruction is possible through accessing model updates \cite{zhu2020deep}.

In this work, we propose an SSL framework, Semi-FedSER, to address the challenge in limited labeled data samples for SER application in the FL setting. Semi-FedSER performs the model training, utilizing both labeled and unlabeled data samples at the local client following FixMatch framework \cite{sohn2020fixmatch}. Furthermore, Semi-FedSER addresses the pseudo-labeling issue mentioned earlier using the idea of multiview pseudo-labeling, and we adopt an efficient yet effective data augmentation technique called Stochastic Feature Augmentation (SFA) \cite{li2021simple}. In addition, we propose to use the past pseudo label information to improve the quality of the pseudo labels generated. To the best of our knowledge, the only similar work that exists in the literature is \cite{tsouvalas2022privacy} for SER application. We show that our Semi-FedSER can generate desired SER performance even when the local label rate $l=20\%$ and data distribution is completely non-iid using IEMOCAP \cite{busso2008iemocap}, and MSP-Improv \cite{busso2016msp} datasets for the experiments.

\vspace{-1.5mm}
\section{SER Data Sets}
\label{sec:data}

In this work, we use two data sets for developing SER models. \autoref{tab:dataset} shows the label distribution of utterances in these corpora used in this work.

\vspace{-1.5mm}
\subsection{IEMOCAP}

The IEMOCAP database \cite{busso2008iemocap} is a multi-modal corpus collected from ten subjects (five male and five female) who target expressing categorical emotions. The data set contains measurements of motion, audio, and video of acted human interactions. The data set contains 10,039 utterances spoken under improvised and scripted conditions. Speakers used material from a fixed script in the scripted condition, while the utterance in the improvised condition is spontaneously spoken. In this work, we focus on the improvised sessions following the suggestions from \cite{zhang2018attention} and our previous works. Due to the data imbalance issue in the IEMOCAP corpus, previous works use the four most frequently occurring emotion labels (neutral, sad, happiness, and anger) for training the SER model \cite{zhang2018attention}. In addition to this, we pick these four emotion classes because most other corpora contain these labels.

\vspace{-1.5mm}
\subsection{MSP-Improv}

The MSP-Improv \cite{busso2016msp} corpus was used to study naturalistic emotions captured from improvised scenarios. The corpus includes audio and video data of utterances spoken in natural condition (2,785 utterances), target condition (652 target utterances in an improvised scenario), improvised condition (4,381 utterances from the remainder of the improvised scenarios), and read speech condition (620 utterances). The data were collected from 12 participants (six male and six female). Similarly to the IEMOCAP data set, we only use the data from the improvised conditions.

\begin{table}
    
    \centering
    \caption{Statistics of emotion labels in two SER data sets.}
    
    \begin{tabular}{p{1.25cm}p{1.25cm}p{1.25cm}p{1.25cm}p{1.25cm}p{1.25cm}}

        \toprule
        \multicolumn{1}{c}{} & 
        \multicolumn{1}{c}{\textbf{Neutral}} & 
        \multicolumn{1}{c}{\textbf{Happy}} &
        \multicolumn{1}{c}{\textbf{Sad}} &
        \multicolumn{1}{c}{\textbf{Angry}} & 
        \multicolumn{1}{c}{\textbf{All}}
        \rule{0pt}{2ex} \\ \midrule

        \multicolumn{1}{l}{\textbf{IEMOCAP}} & 
        \multicolumn{1}{c}{1099} &
        \multicolumn{1}{c}{947} &
        \multicolumn{1}{c}{608} &
        \multicolumn{1}{c}{289} &
        \multicolumn{1}{c}{2943} \rule{0pt}{2ex} \\

        \multicolumn{1}{l}{\textbf{MSP-Improv}} & 
        \multicolumn{1}{c}{2072} &
        \multicolumn{1}{c}{1184} &
        \multicolumn{1}{c}{739} &
        \multicolumn{1}{c}{585} &
        \multicolumn{1}{c}{4580} \rule{0pt}{2ex} \\
        
        \bottomrule
    \end{tabular}
    \vspace{-2mm}
    \label{tab:dataset}
\end{table}

\begin{table}
    \centering
    \caption{Notation used in this paper.}
    \begin{tabular}{p{0.5cm}p{0.5cm}}
        \toprule
        
        \multicolumn{1}{l}{$t$} & 
        \multicolumn{1}{l}{Global training epoch in FL.} \\
        
        \multicolumn{1}{l}{$k$} & 
        \multicolumn{1}{l}{Client index.} \\
        
        \multicolumn{1}{l}{$\mathbf{\theta}^{t}$} & 
        \multicolumn{1}{l}{Global SER model at $t$th global training epoch.} \\
        
        \multicolumn{1}{l}{$\mathbf{\theta}_{k}^{t}$} & 
        \multicolumn{1}{l}{Local SER model updates of $k$th client at $t$th global} \\
        
        \multicolumn{1}{l}{} & 
        \multicolumn{1}{l}{training epoch.} \\

        \multicolumn{1}{l}{$\mathbf{D_{k}^{l}}$} & 
        \multicolumn{1}{l}{Labeled speech data of the kth client.} \\
        
        \multicolumn{1}{l}{$\mathbf{D_{k}^{p}}$} & 
        \multicolumn{1}{l}{Speech data with hard pseudo labels of the kth client.} \\
        
        \multicolumn{1}{l}{$\mathbf{D_{k}^{u}}$} & 
        \multicolumn{1}{l}{Unlabeled speech data of the kth client.} \\

        \multicolumn{1}{l}{$\mathbf{x^{l}}$} & 
        \multicolumn{1}{l}{Labeled speech data.} \\
        
        \multicolumn{1}{l}{$\mathbf{x^{u}}$} & 
        \multicolumn{1}{l}{Unlabeled speech data.} \\
        
        \multicolumn{1}{l}{$\mathbf{x^{p}}$} & 
        \multicolumn{1}{l}{Unlabeled speech data with pseudo labels.} \\
        
        \multicolumn{1}{l}{$\mathbf{\phi(\cdot)}$} & 
        \multicolumn{1}{l}{Weakly-augmentation function.} \\
        
        \multicolumn{1}{l}{$\mathbf{\Phi(\cdot)}$} & 
        \multicolumn{1}{l}{Strongly-augmentation function.} \\
        
        \multicolumn{1}{l}{$\mathbf{y}$} & 
        \multicolumn{1}{l}{Emotion label.} \\
        
        \bottomrule
    \end{tabular}
    \label{tab:notation}
    \vspace{-3.5mm}
\end{table}

\vspace{-0.5mm}
\section{Method}
\label{sec:method}

This section describes the proposed semi-supervised learning framework. To facilitate readability, we summarize the notations adopted in this paper in  \autoref{tab:notation}.

\vspace{-1mm}
\subsection{Model Overview}

In this work, we follow the model framework in FixMatch \cite{sohn2020fixmatch}. We perform two types of augmentations as part of FixMatch: strong and weak, denoted by $\Phi(\cdot)$ and $\phi(\cdot)$, respectively. We generate both weak and strong augmentations through noise addition. Specifically, we adopt the Stochastic Feature Augmentation (SFA) \cite{li2021simple} where given a speech feature $\mathbf{x}$, the weak augmentation is $\phi(\mathbf{x}) = \mathbf{x\odot\mathbf{\alpha} + r},  \; \;\text{where }\mathbf{\alpha} \sim \mathcal{N} (1, \sigma_{1}), \mathbf{r} \sim \mathcal{N} (0, \sigma_{2})$. We choose this method as this efficient yet straightforward feature augmentation technique has shown promising results in model generalization tasks \cite{li2021simple}. We choose a larger $\sigma_{1}$ in $\Phi(\cdot)$ than in $\phi(\cdot)$.

We formulate the learning objectives to include labeled and unlabeled data information. Specifically, our learning function has a supervised loss term $l_{s}$ and an unsupervised loss term $l_{u}$. The supervised loss on labeled data is the standard cross-entropy loss, and given the $k^{th}$ client and its local model with parameter $\theta^{k}$, we aim to minimize the following $l_{s}$ with the weakly-augmented labeled data $\phi(x^{l})$:

\vspace{-1.25mm}
\begin{equation}
    \min \mathcal{L}_{ce}(f(\mathbf{\theta^{k}};\mathbf{\phi(x^{l})}), y)
\end{equation}

When modeling the unlabeled data, our method computes a pseudo label for unlabeled data, which is then used in a standard cross-entropy loss. We obtain the pseudo label $y'$ using the unlabeled weakly-augmented data. We explain our pseudo labeling process in the following subsection. We define the cross-entropy loss against the model's output using the strongly-augmented version of unlabeled data $\mathbf{x^{p}}$:

\vspace{-1.25mm}
\begin{equation}
    \min \mathcal{L}_{ce}(f(\mathbf{\theta^{k}};\mathbf{\Phi(x^{p})}), y')
\end{equation}

Our final goal is to minimize the combined loss below:
\vspace{-1.25mm}
\begin{equation}
    \min l_{s} + l_{u}
\end{equation}

\vspace{-3mm}
\subsection{Pseudo Labeling}

Our Semi-FedSER framework adopts the idea of pseudo labeling proposed in \cite{lin2021semifed}. At each iteration, the model uses predictions from the previous iteration as target classes for unlabeled data samples as if they were true labels. The process goes through a fixed number of iterations until all unlabeled data are labeled. However, the generated pseudo labels can be inaccurate and negatively impact the training results. In order to improve the quality of the generated pseudo labels, \cite{lin2021semifed} proposed to send the aggregated model $\theta^{t}$ and $K$ other local models $\theta_{k}^{t}$ to each client. Then for each local data sample, the local client produces $K+1$ one-hot labels. This approach generates the final pseudo label only if a certain amount of models agrees with the sample label. However, we argue that this method is highly inefficient and unsafe in practice.

\vspace{0.5mm}
\noindent \textbf{Multiview pseudo-labeling (MvPL)} Unlike the pseudo-labeling process in \cite{lin2021semifed}, our framework uses the multiview pseudo-labeling (MvPL) approach proposed in \cite{xiong2021multiview}. The MvPL takes in multiple complementary views of a single unlabeled data point and uses a shared model to generate predictions. Similarly, we decide the final pseudo label by aggregating predictions from all the views. Here, we generate the weakly-augmented views of a speech feature sample using the SFA approach \cite{li2021simple}. Specifically, in our setting, the local client generates $m$ weakly-augmented views using SFA for unlabeled data $\mathbf{x^{u}}$. Suppose there are $m$ predicted class distributions $\mathbf{q_{1}, q_{2}, ..., q_{m}}$, and we can obtain the averaged predicted class distributions as $\mathbf{\bar{q}}$ from all weakly-augmented views. We finally assign the pseudo label $y'$ to $\mathbf{x^{u}}$ if $\max(\mathbf{\bar{q}})$ is above the confidence threshold $\tau$. To further increase the confidence of a prediction, we apply the temperature scaling with a constant scalar temperature $T$ to soften the logits output from the model. Specifically, the probability output for jth class $q_{j}$ is:  

\vspace{-1.5mm}
\begin{equation}
    q_{j} = \frac{e^{z_{j}/T}}{\sum_{i=1}^{C}{e^{z_{i}/T}}}
\end{equation}

Compared to the pseudo-labeling process in \cite{lin2021semifed}, our method is more efficient and does not require sharing local models.

\vspace{1mm}
\noindent \textbf{Pseudo-labeling uncertainty} To further increase the quality of the pseudo labels, we propose to use the idea of uncertainty-aware pseudo label selection process (UPS) \cite{rizve2021defense}. This pseudo-label selection process utilizes both the confidence and uncertainty of a network prediction. We obtain the uncertainty measure $u(\mathbf{q})$ by calculating the standard deviation of $m$ predicted class distributions. We compute the loss for an unlabeled strongly-augmented data $\mathbf{\Phi(x^{u})}$ as:

\vspace{-2.5mm}
\begin{equation}
    \mathbbm{1}(\mathbf{\bar{q}} \geq \tau) \cdot \mathbbm{1}(u(\mathbf{q}) \leq k) \cdot \mathcal{L}_{ce}(f(\mathbf{\theta^{k}};\mathbf{\Phi(x^{u})}), \mathbf{y'})
\end{equation}

Finally, we add data of the pseudo labels to the pseudo-label data and delete them from the unlabeled data. The entire pseudo labeling algorithm is presented in the algorithm \ref{alg:pseudo_labeling}.

\begin{algorithm}
    \caption{MvPL with pseudo-labeling consistency}
    \begin{algorithmic}[1]
        \State{\textbf{Input: }}{$\mathbf{D_{k}^{p}}$, $\mathbf{D_{k}^{u}}$, $\mathbf{\theta_{k}^{t'}}$, $m$, $\tau$, $k$}
        \For{each client $k \in S$ in parallel} 
            \For{$\mathbf{x}^{u} \in \mathbf{D_{k}^{u}}$}
                \State $\mathbf{q} \gets []$
                \For{$i = 1, 2, ..., m$}
                    \State{$\phi_{i}(\mathbf{x}^{u}) = \mathbf{x^{u}\odot\mathbf{\alpha} + r}$ where $\mathbf{\alpha} \sim \mathcal{N} (1, \sigma_{1})$, }
                    
                    \State{$\mathbf{r} \sim \mathcal{N} (0, \sigma_{2})$}
                    
                    \State{append $f(\phi_{i}(\mathbf{x}^{u});\mathbf{\theta}^{t})$ to $\mathbf{q}$}
                    
                \EndFor
            
            \State $\bar{q} \gets {\text{average}(\mathbf{q})}$
            \State $u(\mathbf{q}) \gets {\text{std}(\mathbf{q})}$
            
            \State $y' \gets \argmax{(\bar{q})}$
            
            \If{$\bar{q}\geq\tau$ and $u(\mathbf{q})\leq k$} 
                \State $\mathbf{D_{k}^{p}} \cup \mathbf{x}^{u}$; $\mathbf{D_{k}^{u}} \setminus \mathbf{x}^{u}$
                
            \EndIf
                    
            \EndFor   
        \EndFor
    \end{algorithmic}
    \label{alg:pseudo_labeling}
    
\end{algorithm}

\vspace{-3.75mm}
\subsection{SCAFFOLD}

We often encounter non-IID data at different clients in a typical FL setting. Thus, gradient drifting is a major challenge when training the FL algorithm in a non-IID data setup. To mitigate the impact of the gradient drifting, we propose to implement SCAFFOLD \cite{karimireddy2020scaffold}. SCAFFOLD is a robust algorithm that uses control variates $\mathbf{c}$ (variance reduction) to correct for the 'client-drift' in its local updates. The details of the algorithm proof are in \cite{karimireddy2020scaffold}. Finally, the algorithm \ref{alg:semi-fedser} shows the overall training procedure for our proposed Semi-FedSER pipeline.

\begin{algorithm}
    \caption{Semi-FedSER}
    \begin{algorithmic}[1]
        \State{\textbf{Initialize: }}{$\mathbf{\theta^{0}}$, $\mathbf{c^{0}}$}
        
        \State{\textbf{Server executes:}}
        \For{Each round $t=0,...,T-1$}
            \State{Sample clients $\mathcal{S}\in\{1, 2, ..., K\}$}
            
            \For{Each client $k \in \mathcal{S}$ in parallel}
            
                \State $\mathbf{\theta_{k}^{t}} \gets \mathbf{\theta^{t}}$
                \State $\mathbf{\theta_{k}^{t}}, \mathbf{c_{k}}, \mathbf{\nabla c_{k}} \gets$ ClientLocalTraining($\mathbf{\theta_{k}^{t}}$, $\mathbf{c^{t}}$, $\mathbf{c_{k}}$)
                
            \EndFor
            
            \State $\mathbf{\theta^{t+1}} \gets \frac{1}{|\mathcal{S}|} \sum_{k\in \mathcal{S}} \mathbf{\theta_{k}^{t}}$
            \State $\mathbf{c^{t+1}} \gets \mathbf{c^{t}} + \frac{1}{|\mathcal{S}|} \sum_{k\in\mathcal{S}} \nabla \mathbf{c_{k}}$
        
        \EndFor

        \Function{\textbf{ClientLocalTraining($\mathbf{\theta}$, $\mathbf{c}$, $\mathbf{c'}$)}}{}
            \State $\mathbf{\theta_{old}} \gets \mathbf{\theta}$
            \State $\mathbf{c_{old}} \gets \mathbf{c'}$
            \For{Local epoch $e$ from 0 to $E-1$}
                \For{Iteration $i$ from 0 to $I-1$}
                    \State Sample mini-batch $\mathbf{l}$ from $\mathbf{D_{k}^{l}}$
                    \State Sample mini-batch $\mathbf{p}$ from $\mathbf{D_{k}^{p}}$
                    \State $\mathbf{\theta} \gets \mathbf{\theta}-\eta\nabla_{\theta}(\mathcal{L}_{ce}(f(\mathbf{\theta};\mathbf{\phi(x^{l})}), y) + $
                    
                    \State $\;\;\;\;\;\;\;\;\; \mathcal{L}_{ce}(f(\mathbf{\theta};\mathbf{\Phi(x^{p})}), y')$
                    
                    \State $\mathbf{\theta} \gets \mathbf{\theta} - \eta(\mathbf{c}-\mathbf{c'})$
                    
                \EndFor
                \State \textbf{Execute} Algorithm 1
            \EndFor
            
            \State $\mathbf{c'} \gets \mathbf{c'} - \mathbf{c} + \frac{1}{EI\eta} (\mathbf{\theta_{old}} - \mathbf{\theta})$
            \State ${\textbf{end}}\mathbf{\nabla c} = \mathbf{c'} - \mathbf{c_{old}}$
        
        \State \textbf{return} {$\mathbf{\theta}$, $\mathbf{c'}$, $\mathbf{\nabla c}$}
        \EndFunction
        
    \end{algorithmic}
    \label{alg:semi-fedser}
\end{algorithm}

\vspace{-0.5mm}
\section{Experiments}

In this section, we describe our experimental setup and training details. The implementation of this work is at \href{https://github.com/usc-sail/fed-ser-semi}{https://github.com/usc-sail/fed-ser-semi}.

\vspace{-1mm}
\subsection{Data Preprocessing}

To investigate the effectiveness of the proposed semi-supervised learning framework, we train our SER models on various speech representations. We first evaluate our proposed pipeline using the knowledge-based speech feature set, the Emo-Base feature set, computed from the OpenSMILE toolkit \cite{eyben2010opensmile}. In addition to the knowledge-based speech feature set, we propose to test our framework on SUPERB (Speech Processing Universal PERformance Benchmark) \cite{yang21c_interspeech}, which is designed to provide a standard and comprehensive testbed for pre-trained models on various downstream speech tasks. We compute the deep speech representations from the pre-trained models that are available in SUPERB including APC \cite{chung2019unsupervised}, TERA \cite{liu2021tera}, and DeCoAR 2.0 \cite{ling2020decoar} and DistilHuBERT \cite{chang2021distilhubert}. We choose these three representations as they demonstrated better SER performance in our prior work \cite{feng2021attribute}. Finally, we calculate the global average of the last layer's hidden state as the final feature from the pre-trained model's output. Using the last hidden state is suggested in prior works for downstream tasks \cite{chung2019unsupervised, liu2020nonautoregressive, ling2020decoar, Liu_2020}. Our feature sizes are 988 in Emo-Base; 512 in APC; 768 in Tera, DistilHuBERT, and DeCoAR 2.0. We also apply z-normalization to the speech features within each client.

\begin{table*}

    \centering
    \caption{Prediction results of the SER model trained under different scenarios. The unweighted average recall (UAR) scores of the SER task on each individual data set are reported. The feature sets include Emo-Base obtained from the OpenSMILE toolkit and APC, Tera, DistilHuBERT, and DeCoAR 2.0 computed using pre-trained models. L represents the label rate at the local client.}

    \begin{tabular}{ccccccccccc}
        
        \toprule
        & &
        \multicolumn{3}{c}{\textbf{Fully Supervised}} &
        \multicolumn{2}{c}{\textbf{L=20\%}}  &
        \multicolumn{2}{c}{\textbf{L=40\%}}
        \rule{0pt}{2.25ex} \\ \cmidrule(lr){3-5} \cmidrule(lr){6-7} \cmidrule(lr){8-9}
        
        \multirow{2}{*}{\textbf{Data set}} & 
        \multirow{2}{*}{\textbf{Feature}} &
        \multirow{2}{*}{\textbf{Centralized}} &
        \multirow{2}{*}{\textbf{FedAvg}} &
        \multirow{2}{*}{\textbf{SCAFFOLD}} &
        \multirow{2}{*}{\textbf{Supervised}} & 
        \multicolumn{1}{c}{\textbf{Semi-}} & 
        \multirow{2}{*}{\textbf{Supervised}} & 
        \multicolumn{1}{c}{\textbf{Semi-}} \rule{0pt}{2.25ex} \\

        & & & & & & 
        \multicolumn{1}{c}{\textbf{FedSER}} & & 
        \multicolumn{1}{c}{\textbf{FedSER}}
        
        \rule{0pt}{2.25ex} \\ \cmidrule(lr){1-1} \cmidrule(lr){2-2} \cmidrule(lr){3-5} \cmidrule(lr){6-7} \cmidrule(lr){8-9}
        
        \multirow{4}{*}{\textbf{IEMOCAP}} & 
        \textbf{Emo-Base} & 
        \textbf{61.83\%} & 
        58.01\% & 
        61.34\% & 
        58.49\% & 
        \textbf{60.13\%} & 
        \textbf{59.60\%} & 
        59.51\% 
        \rule{0pt}{1.65ex} \\
        
        & 
        \textbf{APC} & 
        \textbf{66.70\%} & 
        62.07\% & 
        66.21\% & 
        61.21\% & 
        \textbf{63.11\%} & 
        64.98\% & 
        \textbf{65.48\%} 
        \rule{0pt}{1.65ex} \\
        
        & 
        \textbf{DeCoAR 2.0} & 
        \textbf{65.82\%} & 
        62.22\% & 
        65.09\% & 
        58.63\% & 
        \textbf{61.14\%} & 
        62.00\% & 
        \textbf{63.77\% }
        \rule{0pt}{1.65ex} \\
        
        & 
        \textbf{Tera} & 
        \textbf{65.41\%} & 
        60.17\% & 
        64.62\% & 
        56.85\% & 
        \textbf{59.11\%} & 
        63.12\% & 
        \textbf{63.69\%} 
        \rule{0pt}{1.65ex} \\
        
        & 
        \textbf{DistilHuBERT} & 
        \textbf{68.17\%} & 
        62.80\% & 
        67.81\% & 
        61.41\% & 
        \textbf{62.19\%} & 
        62.65\% & 
        \textbf{64.10\%} 
        \rule{0pt}{1.65ex} \\
        
        \cmidrule(lr){1-1} \cmidrule(lr){2-2} \cmidrule(lr){3-5} \cmidrule(lr){6-7} \cmidrule(lr){8-9}
        
        \multirow{4}{*}{\textbf{MSP-Improv}} & 
        \textbf{Emo-Base} & 
        \textbf{48.09\%} & 
        45.85\% & 
        47.73\% & 
        45.56\% & 
        \textbf{46.62\%} & 
        46.20\% & 
        \textbf{47.46\%} 
        \rule{0pt}{1.65ex} \\
        
        & 
        \textbf{APC} & 
        \textbf{53.57\%} & 
        48.33\% & 
        52.91\% & 
        48.55\% & 
        \textbf{50.33\%} & 
        50.71\% & 
        \textbf{50.82\%} 
        \rule{0pt}{1.65ex} \\
        
        & 
        \textbf{DeCoAR 2.0} & 
        \textbf{53.99\%} & 
        49.38\% & 
        53.37\% & 
        47.70\% & 
        \textbf{49.16\%} & 
        51.73\% & 
        \textbf{52.05\%} 
        \rule{0pt}{1.65ex} \\
        
        & 
        \textbf{Tera} & 
        \textbf{53.82\%} & 
        50.38\% & 
        53.81\% & 
        49.26\% & 
        \textbf{50.11\%} & 
        50.91\% & 
        \textbf{51.66\%} 
        \rule{0pt}{1.65ex} \\
        
        & 
        \textbf{DistilHuBERT} & 
        \textbf{55.65\%} & 
        51.22\% & 
        55.37\% & 
        50.51\% & 
        \textbf{50.74\%} & 
        52.01\% & 
        \textbf{53.72\%} 
        \rule{0pt}{1.65ex} \\
        
        \bottomrule

    \end{tabular}
    
    \label{tab:fl_result}
    \vspace{-3.25mm}
\end{table*}

\vspace{-1mm}
\subsection{Non-IID Data setup}

For simulating non-IID distributions over the clients in this experiment, we group the data by each speaker in the IEMOCAP and MSP-Improv data sets and split each speaker's data into 4 different shards. As a result, we produce pathological non-IID data for each partition by taking only 3 unique emotions from each speaker. We regard each partition as a client in federated learning. Consequently, each client in federated learning automatically gets 3 different emotions from one speaker. About 80\% of the speakers are used for training, and the rest are for the test set. 20\% of the data samples in each training client are used as validation, and the rest are training samples. We test our proposed semi-supervised learning pipeline at the labeling rate (training samples) of 20\% and 40\%. We repeat the experiments 5 times with test folds of different speakers in each data set, and we report the average results of the 5-fold experiments.

\vspace{-1mm}
\subsection{Model and training Details}

We use the multilayer perceptron (MLP) as the SER model architecture in this work. The model consists of 2 dense layers with layer sizes of \{256, 128\}. We choose ReLU as the activation function and the dropout rate as 0.2. We implement the SCAFFOLD algorithm in training the SER model. We conduct the experiments on a computer with two NVIDIA GeForce RTX 5000 GPUs to report the performance. We control only 10\% of the clients participating in each global round. 80\% of the data at a client is for local training, and the rest 20\% is for validation. We set the local training batch size as 16. Expressly, we set the learning rate as 0.0001 and the local training epoch as 1 in the FL algorithm. The total global training epochs are 500. We set $\sigma_{2}$ as 0.1 when generating weakly-augment samples and strongly-augment samples. We choose $\sigma_{1}$ as 0.1 and 0.25 in generating weakly-augment samples and strongly-augment samples, respectively. We use a confidence score $\tau=0.5$ in the MvPL algorithm at the beginning of the training and linearly increase it to 0.9 at the 300th global training epoch. We apply a temperature value of 2 to soften the prediction output from weakly-augment samples during the MvPL. We empirically use the number of augmentations $m$ and uncertainty score $k$ as $10$ and $0.005$, respectively. To balance the pseudo label distribution, we add only one data sample with pseudo labels for each class in each training epoch.

\vspace{-1mm}
\section{Results}

\subsection{Compared baselines}
\vspace{-0.5mm}

We compare our semi-supervised learning framework with the following baselines:

\vspace{0.5mm}
\noindent \textbf{Supervised - Federated} Here, we refer to supervised baseline as using only labeled data points at each local client with an assigned labeling rate. We train the supervised Federated baseline using the SCAFFOLD algorithm.

\vspace{0.5mm}
\noindent \textbf{Fully Supervised - Federated} We train a fully supervised Federated baseline (use all the original labels of the training data) to measure the ideal accuracy we would obtain under federated learning settings. We train the fully supervised Federated baseline using both FedAvg and SCAFFOLD.

\vspace{0.5mm}
\noindent \textbf{Fully Supervised - Centralized} Finally, we train a fully supervised centralized baseline to measure the upper bound SER model performance in different data sets.

\vspace{-1mm}
\subsection{SER performance}

\noindent \textbf{Baselines} The emotion prediction results of FL training on different data sets are shown in \autoref{tab:fl_result}. We report the SER prediction results in unweighted average recall (UAR) scores. We observe that our fully-supervised federated SER performance using the SCAFFOLD algorithm yields similar performance to prior works in \cite{satt2017efficient, ramet2018context}. Moreover, we can identify that the performance of the fully supervised SER model trained in a centralized setup is slightly higher than the fully supervised SER model trained using the SCAFFOLD algorithm. Meanwhile, we find that the SCAFFOLD algorithm performs consistently better than the FedAvg algorithm in training the fully supervised SER model. These results suggest that the SCAFFOLD algorithm can significantly increase the model performance in the federated learning setup.

\noindent \textbf{Semi-FedSER} When the label rate of the local client is at 20\%, we can observe that the SER performance of the federated supervised model drops 5-10\% compared to the fully federated supervised model on the different data sets and using different speech features. However, we can observe that SER models trained using our proposed Semi-FedSER framework generate similar UAR scores to fully federated supervised models. For example, the UAR score of the fully supervised model using FL is 66.70\% in the IEMOCAP data set using the APC speech feature. In contrast, the SER model trained using our proposed Semi-FedSER framework on the same speech feature produces a similar UAR score of 63.11\%. We can further observe that the performance of the Semi-FedSER framework is at the same level as the fully supervised FL model across all testing scenarios when the label rate is 40\%. These comparisons demonstrate that the proposed Semi-FedSER can achieve desirable performance with few on-device labeled samples.

\vspace{-1mm}
\section{Conclusions}

In this work, we propose a novel Semi-FedSER framework to address the challenges in limited labeled data sample settings for speech emotion recognition in FL setting. Semi-FedSER leverages both labeled and unlabeled data samples at the local client combined with pseudo-labeling. The pseudo-labeling approach in Semi-FedSER is based on the multiview pseudo-labeling. We further improve the quality of the pseudo-labeling by incorporating the pseudo-labeling consistency measurement. We also implement the SCAFFOLD algorithm to address the non-IID data distribution issue in the FL setting. Our results show that the proposed Semi-FedSER framework effectively generates desired SER predictions even when the local label rate $l=20\%$. In the future, we plan to extend our current work to multi-modal data scenarios where not every modality is presented with labels. We also aim to design an automatic scheme for deciding the threshold $\tau$ in generating pseudo labels.


\bibliographystyle{IEEEtran}

\bibliography{mybib}

\end{document}